\documentclass{emulateapj}

\shorttitle{Progenitors of super-Chandrasekhar mass SNe Ia }
\shortauthors{Chen \& Li}

\begin{document}


\title{On the progenitors of super-Chandrasekhar mass type Ia supernovae}


\author{Wen-Cong Chen$^{1}$   and Xiang-Dong Li$^{2}$}
\affil{$^1$Department of Physics, Shangqiu Normal University,
Shangqiu 476000, China; chenwc@nju.edu.cn\\
$^2$Department of Astronomy, Nanjing University,
    Nanjing 210093, China; lixd@nju.edu.cn}



\begin{abstract}
Type Ia supernovae (SNe Ia) can be used  as the standard
candlelight to determine the cosmological distances because they
are thought to have a uniform fuel amount. Recent observations of
several overluminous SNe Ia suggest that the white dwarf masses at
supernova explosion may significantly exceed the canonical
Chandrasekhar mass limit. These massive white dwarfs may be
supported by rapid differential rotation.  Based on single
degenerate model and the assumption that the white dwarf would
differentially rotate when the accretion rate
$\dot{M}>3\times10^{-7}M_{\odot}\rm yr^{-1}$, we have calculated
the evolutions of close binaries consisting of a white dwarf and a
normal companion. To include the effect of rotation, we introduce
an effective mass $M_{\rm eff}$ for white dwarfs. For the donor
stars with two different metallicities $Z=0.02$ and 0.001, we
present the distribution of the initial donor star masses and the
orbital periods of the progenitors of super-Chandrasekhar mass SNe
Ia. The calculation results indicate that, for an initial massive
white dwarf of $1.2M_{\odot}$, a considerable fraction of SNe Ia
may result from super-Chandresekhar mass white dwarfs, but very
massive ($>1.7 M_{\sun}$) white dwarfs are difficult to form, and
none of them could be found in old populations. However,
super-Chandrasekhar mass SNe Ia are very rare when the initial
mass of white dwarfs is $1.0M_{\odot}$. Additionally, SNe Ia in
low metallicity environment are more likely to be homogeneous.
\end{abstract}


\section{Introduction}
Type Ia supernovae (SNe Ia) are generally believed to be
thermonuclear explosions of accreting carbon-oxygen white dwarfs
\citep{hoy60} when their masses reach the chandrasekhar mass of
$\sim 1.4 M_{\odot}$ \citep{chan31}. Due to this uniform
progenitor mass \citep{mazz07}, most SNe Ia present a good
correlation between the peak brightness and the width of the light
curve \citep{phil93}, so it is possible to use them as the
standard candlelight to determine the cosmological distances
\citep{ries98,ries04,perl99}.

There exist double degenerate model \citep{iben84,webb84} and
single degenerate model \citep{whel73,nomo82} for the progenitors
of SNe Ia \citep[for a review see][]{bran95}. Based on the single
degenerate model, the evolutions of binary systems consisting of
an accreting white dwarf have been widely explored by many authors
\citep{hach96,li97,hach99a,hach99b,yoon03,han04,chen07,hach08,meng08,xu08,han08,wang09}.

The supernova SNLS-03D3bb (2003fg) was discovered
on 24 April 2003, and observed to be 2.2 times
overluminous than a normal SN Ia \citep{asti06}. \citet{howe06}
inferred the mass of $^{56}$Ni $\sim 1.3 M_{\odot}$, which
indicates that the mass of the white dwarf at the moment of
explosion is $\sim 2.1 M_{\odot}$\footnote{
\citet{hill07} suggested an alternative model to account for the
high brightness of SNLS-03D3bb, in which an off-center ignition of
nuclear burning in a Chandrasekhar mass white dwarf occurs and
drives a lop-sided explosion}. \citet{wang08} also suggested
white dwarfs with super-Chandrasekhar masses as the
progenitors of some rare nickel-rich SNe Ia ($M_{\rm
Ni}> 0.8 M_{\odot}$). These massive white dwarfs may be supported by  rapid
rotation \citep{yoon05}, or origin from the merger of
two massive white dwarfs \citep{tutu94,howe01}. More recently,
two other possibly overluminous SNe Ia (SN 2006gz, \citet{hick07} and
SN 2007if, \citet{yuan07}) were reported.

It is interesting to explore the properties of the progenitors of
overluminous SNe Ia like SNLS-03D3bb in the single degenerate
model. Recently, some authors have taken account of the influence
of rotation on the accreting white dwarf in their works
\citep{ueni03,saio04,yoon04a,yoon05}. \citet{yoon04a} found that a
white dwarf with accretion rate of $\ga 3\times10^{-7}M_{\odot}
\rm yr^{-1}$ may rotate differentially. The maximum equilibrium
mass for a secularly stable white dwarf with differential rotation
is given by \citep{shap83}
\begin{equation}
M_{\rm max}\approx 2.5(\frac{2}{\mu_{\rm e}})^{2}M_{\odot},
\end{equation}
where $\mu_{\rm e}$ is the mean molecular weight per electron.
This value is in line with the results derived by \citet{Duri75}
and \cite{Duri81} via detailed numerical calculations.

Considering the angular momentum transfer by mass accretion,
\citet{yoon04b} simulated the evolution of helium-accreting CO
white dwarfs for different mass accretion rate, and found that the
rotation can help white dwarfs grow in mass by stabilizing helium
shell burning. Based on the rigidly rotating progenitor model,
\citet{Domi06} computed the evolution of a rotating CO white dwarf
accreting  CO-rich matter. Their results show that more massive
progenitors result in higher $^{56}$Ni mass and explosive
luminosity, and more massive white dwarfs at explosion.

Assuming that overluminous SNe Ia originate from binary systems consisting of
an accreting white dwarf with super-Chandrasekhar mass, in this
paper we investigate the distribution of the initial donor
star mass and orbital period of the progenitor binaries.
In section 2, we describe the input
physics in stellar and binary evolution calculations. Numerically
calculated results for the evolutionary sequences of white dwarf
binaries are presented in section 3. In section 4 we summarize the results
with a brief discussion on the limitation of this study.

\section[]{Input Physics}
In this work, we study the evolution of close binaries consisting
a CO white dwarf (of mass $M_{\rm wd}$) and a normal
companion (of mass $M_{\rm d}$) with population I
and II metallicities ($Z=0.02$ and 0.001, respectively). The white dwarf binary
results from a binary consisting of two main-sequence stars, in
which the more massive star evolves more rapidly, and becomes  the
CO white dwarf. With nuclear evolution, the secondary
star starts to fill its Roche lobe and transfer hydrogen-rich
material to the white dwarf. Our calculations begin at this stage.

The accreted material on the white dwarf will be heated and
compressed, finally leading to nuclear burning. If the burning
process is unstable, part of the mass is ejected from the white
dwarf. The accumulated efficiencies for hydrogen and helium
burning are denoted to be $\alpha_{\rm H}$ and $\alpha_{\rm He}$,
respectively.
Although calculations of $\alpha_{\rm H}$
and $\alpha_{\rm He}$ have been done by, e.g., \citet{kove94}
and \citet{kato04},  they are limited to non-rotating white dwarfs,
and in principle, inadequate to determine the final
mass of H accreting rotating white dwarfs.
To explore the effect of stellar rotation on the thermal response of accreting
CO white dwarfs, we include the lifting effect in the hydrostatic equilibrium
according to the prescriptions in \citet{Domi96},
and introduce an effective mass $M_{\rm
eff}$ of the white dwarf by taking account of the centrifugal force.
We then adopt the corresponding values of $\alpha_{\rm H}$ and $\alpha_{\rm He}$
for this effective mass from
the formulae in \citet{han04} and \citet{kato04}. The method can
be described as follows. We divide the surface of the white dwarf into three zones
with different ranges of the polar angle:
the equatorial zone ($\theta=60^{\circ}-120^{\circ}$),
the middle zone ($\theta=30^{\circ}-60^{\circ} $ and
$120^{\circ}-150^{\circ} $), and the polar zone
($\theta=0^{\circ}-30^{\circ}$ and
$150^{\circ}-180^{\circ} $). Assume rigid
body rotation and neglect any deformation of the white dwarf,
the effective mass $M_{\rm
eff}$ of the white dwarf in the zone with polar angles between
$\theta_{1}$ and $\theta_{2}$ satisfies
\begin{equation}
\frac{GM_{\rm eff}}{R^{2}}= \frac{GM_{\rm wd}}{R^{2}} -
\frac{\int^{\theta_{2}}_{\theta_{1}}2\pi \omega
^{2}R^{3}\rm{sin}^{3}\theta
\rm{d}\theta}{\int^{\theta_{2}}_{\theta_{1}}2\pi
R^{2}\rm{sin}\theta \rm{d}\theta},
\end{equation}
where $R$ and $\omega$ are the radius and angular velocity of the white dwarf,
respectively. In Eq.~(2), the radial
component of the centrifugal force is averaged by an area-weighted
mean.

We further assume that each zone accretes the transferred material
at a rate proportional to its area, i.e. its accretion fraction
$f_{\rm i}=\int^{\theta_{2}}_{\theta_{1}}2\pi R^{2}\rm{sin}\theta
\rm{d}\theta/4\pi \emph {R}^{2}$, and $f_{\rm i}=0.5$, 0.366, and
0.134 for the equatorial zone, middle zone, and polar zone,
respectively. Replacing $M_{\rm wd}$ with $M_{\rm eff}$, we can
simply obtain the H and He accumulated efficiencies ($\alpha_{\rm
H,i}$ and $\alpha_{\rm He,i}$) for different zones on the surface
of the white dwarf.
Summarize the above prescriptions, the mass growth
rate of the white dwarf can then be written as
\begin{equation}
\dot{M}_{\rm wd}= \sum \alpha_{\rm H,i}\alpha_{\rm He,i}f_{\rm
i}\dot{M_{\rm d}}.
\end{equation}
where $\dot{M_{\rm d}}$ is the mass transfer rate from the donor star.

We also calculate the spin evolution of  the white
dwarf, neglecting the possible interaction between the magnetic
field of the white dwarf and the accretion disk.
The rotation of the white dwarf may be related to the
rotation of its progenitor star \citep{hege00,maed00}, but is more
likely to be attained during mass transfer, as the transferred
material from the donor star carries a large amount of angular
momentum, which can cause spin-up of the white dwarf
\citep{Duri77,ritt85,nara89,lang00}. Generally the white dwarf will spin up along
with accretion from a disk \citep{Duri77,ritt85}. When the rotation of
the white dwarf is close to break-up,  numerical calculations showed that angular momentum will be
transferred from the white dwarf into the accretion disk, and the rotation
velocity will roughly keep constant \citep{pacz91,poph91}.
Hence we set the specific
angular momentum of the effective accreted matter to be
\begin{equation}
j_{\rm acc}= \left\{\begin{array}{l@{\quad}l}  \sqrt{GM_{\rm
wd}R},&v<0.9v_{\rm K}, \strut\\
0\quad , &  v\geq0.9v_{\rm K} \strut\\
\end{array}\right.
\end{equation}
where $v=\omega R$, and $v_{\rm K}=\sqrt{GM_{\rm wd}/R}$ are the
rotation velocity and the Keplerian velocity at the white dwarf's
equator, respectively. In our calculations, the initial surface
velocity at the white dwarf's equator is taken to be $10 \rm
km\,s^{-1}$, and the white dwarf radius changes with $R\propto
M_{\rm wd}^{-1/3}$.

The losses of mass and orbital angular momentum play an important
role in the mass transfer and orbital evolution of close binary
systems. The mass loss rate of the binary system is
$\dot{M}=(1-\sum \alpha_{\rm H,i}\alpha_{\rm He,i}f_{\rm
i})\dot{M_{\rm d}}$, and we assume that this mass is ejected in the
vicinity of the white dwarf in the form of isotropic winds or
outflows, taking away the specific orbital angular momentum of the
white dwarf. The rate of orbital angular momentum loss through mass loss is given by
\begin{equation}
\dot{J}_{\rm is}=(1-\sum \alpha_{\rm H,i}\alpha_{\rm He,i}f_{\rm
i})\frac{\dot{M_{\rm d}}M_{\rm d}}{MM_{\rm wd}}J,
\end{equation}
where $J=(M_{\rm d}M_{\rm wd}/M)a^{2}\Omega$, $M=M_{\rm d}+M_{\rm
wd}$, and $\Omega$ are the  orbital angular momentum,
total mass, and orbital angular velocity of the binary system,
respectively.

 With numerical simulations,
\citet{yoon04a} found that, accreting at a rate
$\ga3\times10^{-7}M_{\odot}\rm yr^{-1}$, a white dwarf would
rotate differentially, and there would not be the central carbon
ignition even if its mass exceed $1.4M_{\odot}$. As a result of
the differential rotation, a white dwarf with a
super-Chandrasekhar mass may exist. Accordingly, we  assume that a
SN Ia occurs when $M_{\rm wd}\geq 1.4 M_{\odot}$ and
$\dot{M}<3\times10^{-7}M_{\odot}\rm yr^{-1}$, so that there is no
differential rotation to support the massive white dwarf. In the
case of $M_{\rm wd}> 1.4M_{\odot}$ and
$\dot{M}>3\times10^{-7}M_{\odot}\rm yr^{-1}$, we let the white
dwarf increase mass up to $M_{\rm max}$.

\section[]{Results}
We have calculated the evolution of white dwarf binaries adopting
an updated version of the stellar evolution code developed by
\citet{eggl71,eggl72} \citep[see also][]{han94,pols95}.  The
stellar OPAL opacities are from \citet{roge92}, and \citet{alex94}
for a low temperatures. In the calculations we take the ratio of
the mixing length to the pressure scale height to be 2.0. The
evolutionary results are determined by three parameters of white
dwarf binaries: the initial mass of the white dwarf $M_{\rm wd,i}$
(taken to be $1.2M_{\odot}$ and $1.0M_{\odot}$), the initial mass
of the donor star $M_{\rm d,i}$, and the initial orbital period of
binary systems $P_{\rm orb,i}$.

We first present an example of the evolutionary sequences for a
binary system with $M_{\rm wd,i}=1.2M_{\odot}$, $M_{\rm
d,i}=2.5M_{\odot}$, and $P_{\rm orb,i}=1.0$ day, in which the
donor star has a solar composition ($Y=0.28$, $Z=0.02$). In Figure
1 we plot the evolution of the mass transfer rate and the donor
star mass with time in the left panel, the evolution of the
orbital period and the white dwarf mass in the middle panel, and
the evolution of rotation velocity at the white dwarf's equator in
the right panel. The donor star fills its Roche lobe when its age
is $\sim 3.84\times 10^{8}$ yr. Because of a relatively high
initial mass ratio, the mass transfer occurs on a thermal
timescale at a high rate of $\sim 10^{-7}-10^{-6}M_{\odot}\rm
yr^{-1}$, but remains stable due to strong isotropic wind from the
white dwarf. After $\sim 1.5$ Myr mass transfer, the white dwarf
grows to $ 1.75M_{\odot}$, and trigger a type Ia supernova when
the mass transfer rate $\dot{M}$ declines to be
$3\times10^{-7}M_{\odot}\rm yr^{-1}$. The orbital period firstly
decreases to 0.7 day, because material is transferred from the
more massive donor star to the less massive white dwarf, and then
increases when the white dwarf mass grows and exceeds the donor
star mass. With gaining spin angular momentum from the accretion
material, the rotation velocity of the white dwarf gradually
increase, and near the break-up velocity.

To investigate the distribution of the initial donor star mass and
orbital period for the progenitor systems of super-Chandrasekhar
mass SNe Ia, we have calculated the evolutions of a large number
of white dwarf binaries with different values of $M_{\rm d,i}$ and
$P_{\rm orb,i}$ when $M_{\rm wd,i}=1.2M_{\odot}$. Figure 2
summarizes our calculated results in the $M_{\rm d,i}-P_{\rm
orb,i}$ plane with $Z=0.02$ (left panel) and 0.001 (right panel),
respectively. In the case of $Z=0.02$, the regions enclosed by the
solid, dashed, and dotted curves denote the distribution areas of
initial white dwarf binaries with $M_{\rm SN}\ge 1.7M_{\odot}$,
$\ge 1.6M_{\odot}$, and $\ge1.4 M_{\odot}$, respectively. Beyond
these areas, SNe Ia cannot occur due to either a low mass
accumulation efficiency of the white dwarf or unstable mass
transfer. In the case of $Z=0.001$, the solid, dashed, and dotted
curves correspond to the boundaries of distribution area for the
progenitors with $M_{\rm SN}\ge 1.6M_{\odot}$,  $\ge1.5
M_{\odot}$, and $\ge1.4 M_{\odot}$, respectively, and we cannot
find that a white dwarf mass exceed $1.7 M_{\odot}$. When the
metallicity $Z$ changes from 0.02 to 0.001, the occupied region by
the progenitors of SNe Ia has a tendency to move downward  in the
$M_{\rm d,i}-P_{\rm orb,i}$ diagram, i.e., the progenitors with
higher $Z$  tend to have a more massive donor star, in agreement
with \citet{meng08}.  However, it is difficult to produce a
super-Chandrasekhar mass SN Ia when the donor mass is low as $1.0
M_{\odot}$. Figure 3 presents the progenitor distribution of SNe
Ia in the $M_{\rm d,i}-P_{\rm orb,i}$ plane when $M_{\rm
wd,i}=1.0M_{\odot}$. For $Z=0.02$, the maximum explosion mass of
white dwarfs is always less than $1.5 M_{\odot}$.

Figure 4 shows the distribution of $M_{\rm d,f}$ and $P_{\rm
orb,f}$ at the moment of SN explosion. The open triangles, open
circles, and solid stars denote systems with $1.6M_{\odot}> M_{\rm
SN}\ge 1.4M_{\odot}$, $1.7M_{\odot}> M_{\rm SN}\ge1.6M_{\odot}$,
and $M_{\rm SN}\ge 1.7M_{\odot}$  in the left panel ($Z=0.02$),
and $1.5M_{\odot}> M_{\rm SN}\ge 1.4M_{\odot}$, $1.6M_{\odot}>
M_{\rm SN}\ge 1.5M_{\odot}$, and $M_{\rm SN}\ge 1.6M_{\odot}$  in
the right panel ($Z=0.001$), respectively. For  $Z=0.02$ and
$M_{\rm SN}\ge 1.7M_{\odot}$, the companion masses $\sim 1.3-
1.8M_{\odot}$ and orbital periods $\sim 0.4-2.0$ d, while in the
case of $Z=0.001$ and $M_{\rm SN}\ge 1.6M_{\odot}$, the companion
masses $\sim 0.5- 1.5M_{\odot}$ and orbital periods $\sim 0.4-6.0$
d. In Figure 5, we plot the distribution of the companion stars at
the moment of SNe Ia in the H-R diagram. They have luminosities in
the range of $\sim1.0-10L_{\odot}$ and effective temperatures in
the range of $\sim 5000-6000$ K. These properties may be compared
with and testified by future optical observations of SN Ia
remnants.

Certainly our approach in estimating the mass accumulation
efficiency on rotating white dwarfs is very rough and simplified.
We only consider the effect of centrifugal force and neglect the
thermal and chemical evolution on the surface of rapid rotating
white dwarfs, which is very complicated. In our calculations we
assumed a rigid rotation instead of a differential rotation, which
only increase slightly the Chandrasekhar limit to be
$1.48M_{\odot}$ \citep{yoon04a}. Additionally, the white dwarf
will be deformed during spin-up, not obeying spherical symmetry as
we assumed. Nevertheless, Our results show that the mass
accumulation efficiency of white dwarfs decrease with rotation.
This is at least qualitatively consistent with \citet{pier03a}, in
which the authors compared rigidly rotating white dwarfs with
non-rotating objects for CO-accretion at various accretion rate,
and found that an increase in the angular velocity of the white
dwarf always leads to a decrease in the value of the accretion
rate below which central carbon ignition will occur. The reason is
that including rotation cause the accreting star to be both less
dense and cooler, resulting in a larger thermal diffusion
timescale, while the effect of compression induced by the
accretion process is only slightly modified. This implies that our
treatment might not be far from real situations.

The influence of metallicities on the mass accumulation efficiency
can be found in Figure 6, which shows the fraction of mass growth
in the transferred mass
\begin{equation}
\eta=\frac{M_{\rm wd,f}-M_{\rm wd,i}}{M_{\rm d,i}-M_{\rm d,f}}
\end{equation}
against the initial orbital period $P_{\rm orb,s}$ when the initial mass of the
donor star $M_{\rm d.i}=2.5M_{\odot}$. It is clearly seen that, for the
same donor star, white dwarfs accompanied by a
Population I star have a higher mass growth rate.

\section[]{Discussion and Summary}

Recently discovered overluminous SNe Ia suggested that they might
have originated from super-Chandrasekhar mass white dwarfs
\citep{howe06,hick07,yuan07}. Based on the single degenerate model
and the suggestion that massive white dwarfs might be supported by
rapid differential rotation when accretion rate $\dot{M}\ga
3\times 10^{-7}M_{\odot}\rm yr^{-1}$ \citep{Duri75,shap83,yoon05},
we have performed numerical calculations of the evolution of white
dwarf binaries to investigate the properties of the progenitor of
super-Chandrasekhar mass SNe Ia. The results can be summarized as
follows.

(1) For white dwarfs with an initial mass of $\la 1.0 M_{\sun}$,
the explosion masses of SNe Ia are nearly uniform, i.e.
super-Chandrasekhar mass SNe Ia are difficult to produce. This is
consistent with the rareness of super-Chandrasekhar mass SNe Ia in
observations.

(2) When $M_{\rm wd,i}=1.2 M_{\sun}$, depending on the
evolutionary paths the masses ($M_{\rm SN}$) of exploding white
dwarfs range from $1.4 M_{\sun}$ to $1.76 M_{\sun}$. A
considerable fraction of SNe Ia are of super-Chandrasekhar mass,
suggesting a diversity in the brightness of SNe Ia. However, in
most cases the final masses of the white dwarfs are not
significantly exceeding $1.4\,M_{\sun}$, and it is very difficult
to produce a SN Ia with $M_{\rm SN}\ga1.8\,M_{\sun}$. Thus our
model cannot reproduce an overluminous SN Ia with $M_{\rm SN}\ga
2.0 M_{\sun}$ like SNLS-03D3bb \citep{howe06}\footnote{It is
controversial if the progenitor of SN 2003fg is a
super-Chandrasekhar mass white dwarf. Based on the low velocity
and short time-scale seen in SN 2003fg, \citet{maed08b} suggested
that its ejecta mass is smaller than $1.4M_{\odot}$.}.

(3) Progenitors of super-Chandrasekhar mass ($M_{\rm
SN}\ge1.6M_{\odot}$) SNe Ia can be constrained to be white dwarf
(with initial mass of $1.2M_{\odot}$) binaries with an initial
donor star mass of $M_{\rm d,i}\sim 2.2-3.3M_{\odot}$ and an
initial orbital period of $P_{\rm orb,i}\sim 0.5-4.0$ d when
$Z=0.02$, and $M_{\rm d,i}\sim 1.7-2.7M_{\odot}$ and $P_{\rm
orb,i}\sim 0.5-3.5$ d when $Z=0.001$. Interestingly, it is not
those with most massive donors that are more likely to evolve to
super-Chandrasekhar mass SNe Ia, since the mass transfer rates in
these systems usually decline rapidly below
$3\times10^{-7}M_{\odot}\rm yr^{-1}$, so that the white dwarf does
not have sufficient time to accrete. The time delay between the
formation of the progenitor systems and the explosions of SNe Ia
is $\la 1$ Gyr, in agreement with the suggestion that
super-Chandrasekhar mass SNe Ia should be more likely to exist in
a young stellar population \citep{howe06}, but they seem not to
belong the youngest population. Especially one would not expect
overluminous SNe Ia in early type galaxies.

(4) Metallicities have important influence on the production of
super-Chandrasekhar mass SNe Ia. Systems with Population II donor
stars are less likely to be the progenitor of SN Ia with a
super-Chandrasekhar mass of $\ga1.7M_{\odot}$, suggesting that SNe
Ia at relatively high redshifts might be more homogeneous than nearby
ones.

Perhaps the biggest issue in this work is how super-Chandrasekhar
mass white dwarfs can be formed and related to overluminous SNe
Ia.  The nature of the overluminous SNe Ia is not yet well
understood. For example, Subaru and Keck optical spectroscopic and
photometric observations of SN Ia 2006gz show that,  the late-time
behavior of this SN is distinctly different from that of normal
SNe Ia, but the peculiar features found at late times (the SN is
faint and it lacks [Fe II] and [Fe III] emission) are not readily
connected to a large amount of $^{56}$Ni \citep{maed08a}. Even if
super-Chandrasekhar mass white dwarfs do exist in the universe,
their formation path is poorly known. In principle, both merge of
a close binary system of two massive white dwarfs and accretion
onto a normal white dwarf may form a rapidly spinning white dwarf
that considerably exceeds the Chandrasekhar limit,  provided that
they rotate differentially. In the latter case, under what
circumstances can a differentially rotating white dwarf be
sustained has not been well addressed. We have adopted the simple
criterion of the mass accretion rate, but the real situation
should be sensitively dependent on the efficiency of mass and
angular momentum gain and loss during mass transfer, which may be
also related to the magnetic field strength of the white dwarfs.

\acknowledgments { We are grateful to the anonymous referee for
his/her constructive suggestion improving this manuscript, and
Zhan-Wen Han for helpful discussions. This work was partly
supported by the National Science Foundation of China (under grant
number 10873011 and 10873008), the National Basic Research Program
of China (973 Program 2009CB824800), and Program for Science \&
Technology Innovation Talents in Universities of Henan Province,
China.}

\clearpage
\begin{figure}
\epsscale{.3} \plotone{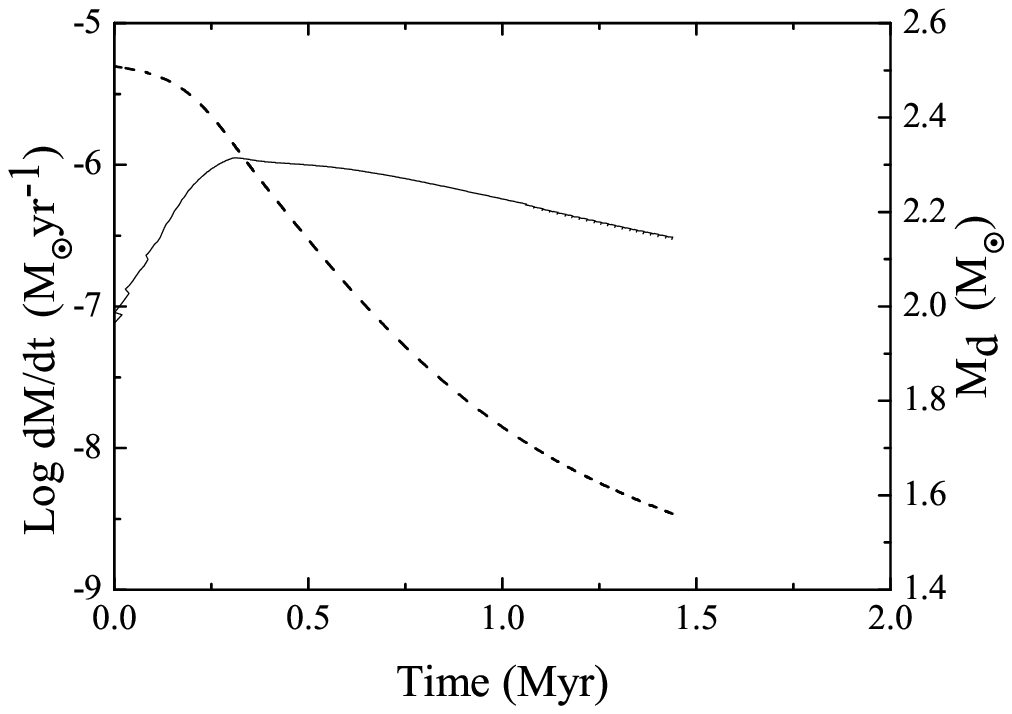}\plotone{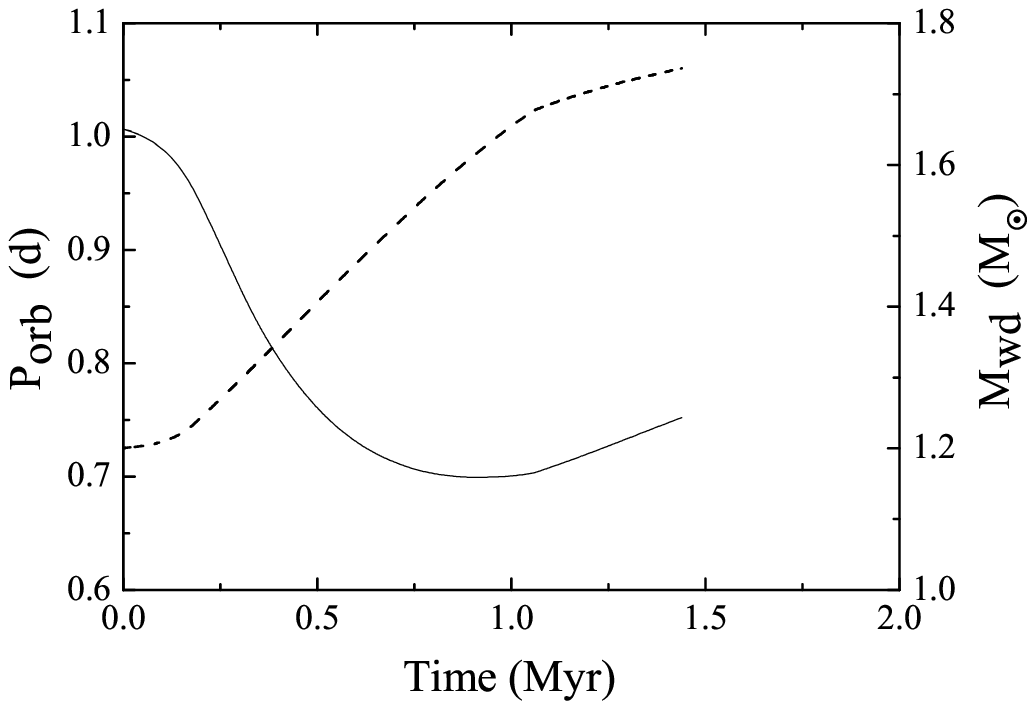}
\plotone{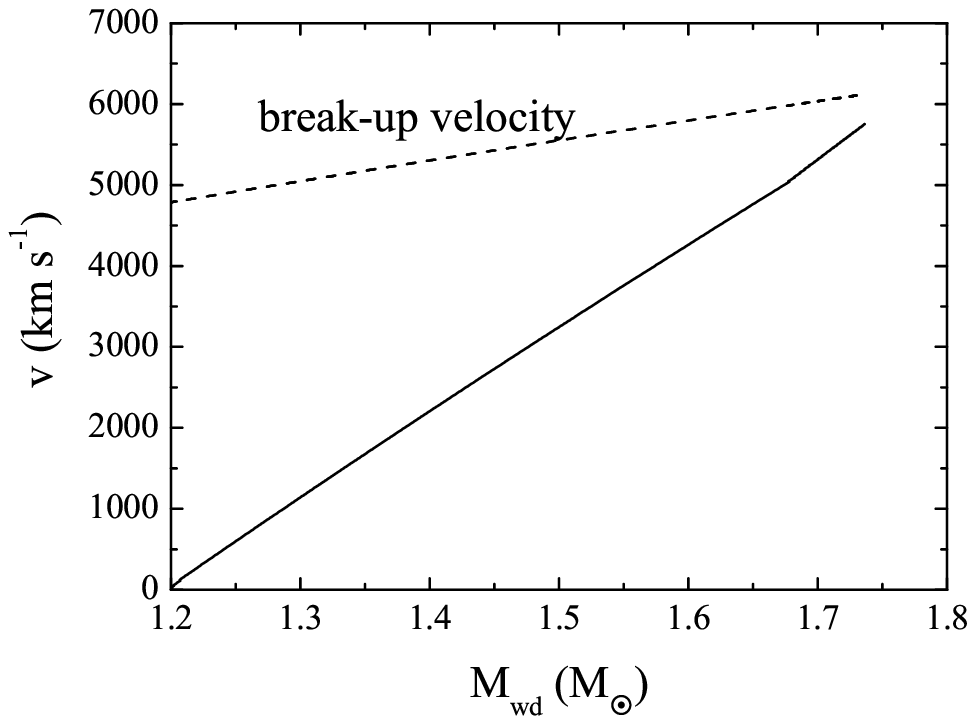}\caption{Evolution of a white dwarf binary with
$M_{\rm wd,i}=1.2M_{\odot}, M_{\rm d,i}=2.5M_{\odot}$ and $P_{\rm
orb,i}=1.0\rm day$. The solid and dashed curves denote the
evolutionary tracks of the mass transfer rate and the donor star
mass in the left panel, the orbital period and the white dwarf
mass in the middle panel, and the rotation velocity and break-up
velocity at the white dwarf's equator, respectively.}
\label{FigVibStab}
\end{figure}

\epsscale{1.}
\begin{figure}
\plottwo{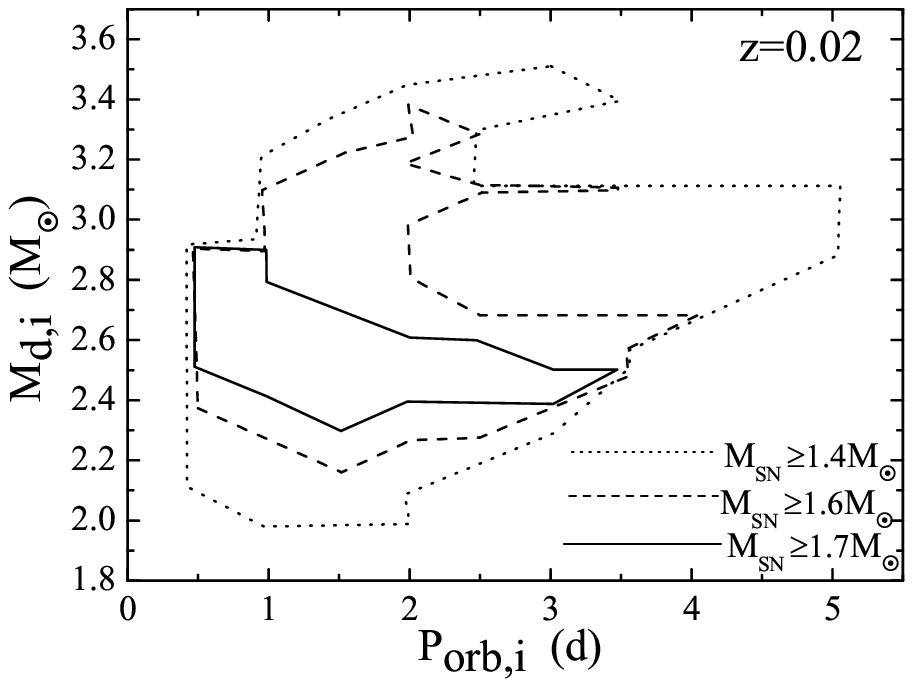}{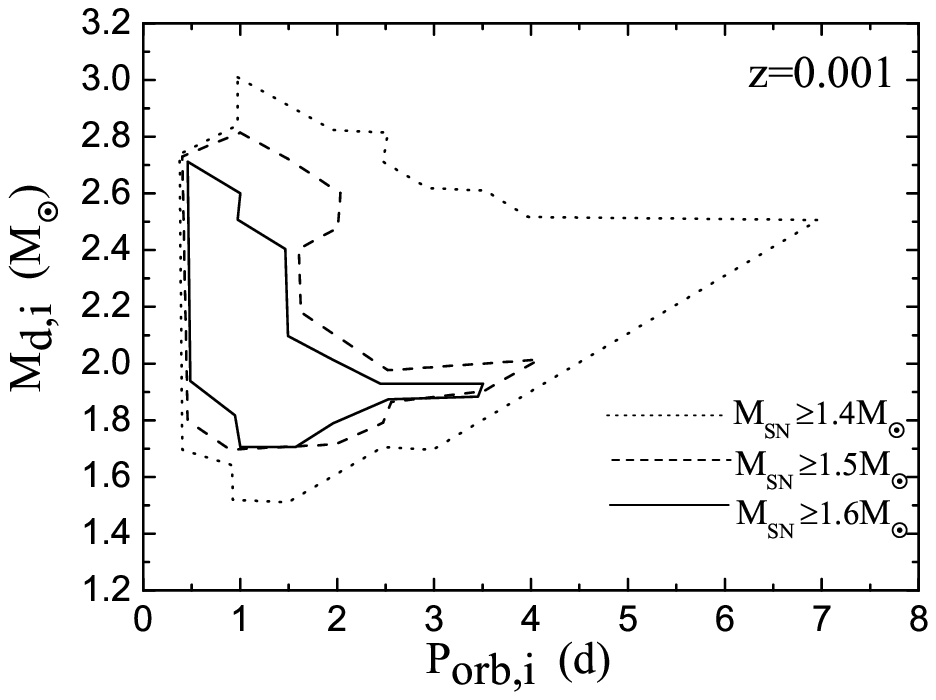} \caption{Distribution of the initial
orbital periods $P_{\rm orb,i}$ and the initial donor star masses
$M_{\rm d,i}$  of the progenitor systems of super-Chandrasekhar
mass SNe Ia when $M_{\rm wd,i}=1.2M_{\odot}$. The left and right
panels are for $Z=0.02$ and 0.001, respectively.}
\label{FigVibStab}
\end{figure}

\begin{figure}
\plottwo{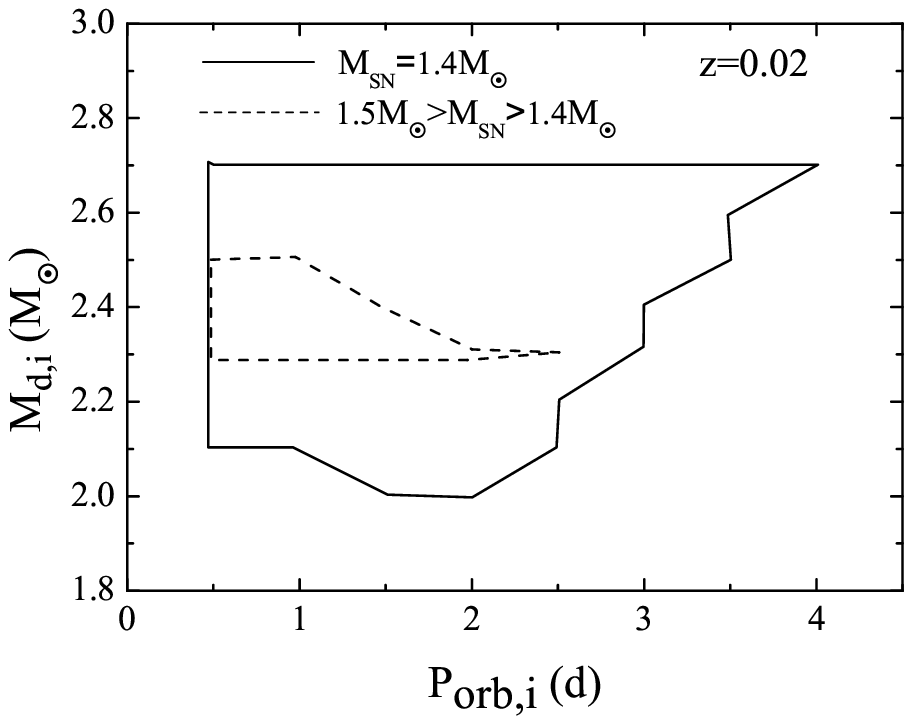}{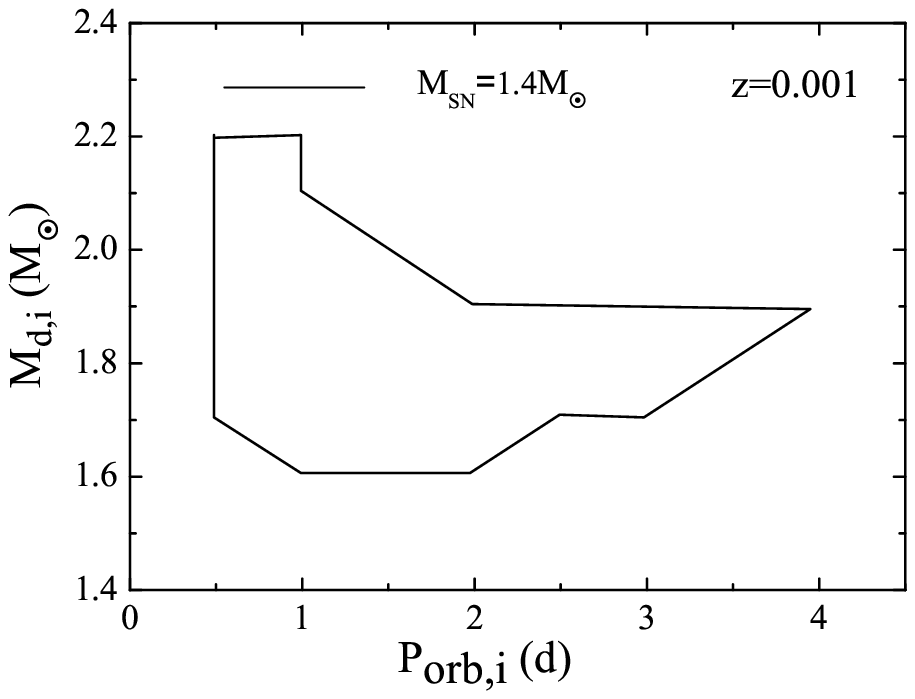} \caption{Distribution of the initial
orbital periods $P_{\rm orb,i}$ and the initial donor star masses
$M_{\rm d,i}$  of the progenitor systems of SNe Ia when $M_{\rm
wd,i}=1.0M_{\odot}$. The left and right panels are for $Z=0.02$
and 0.001, respectively.} \label{FigVibStab}
\end{figure}

\begin{figure}
\plottwo{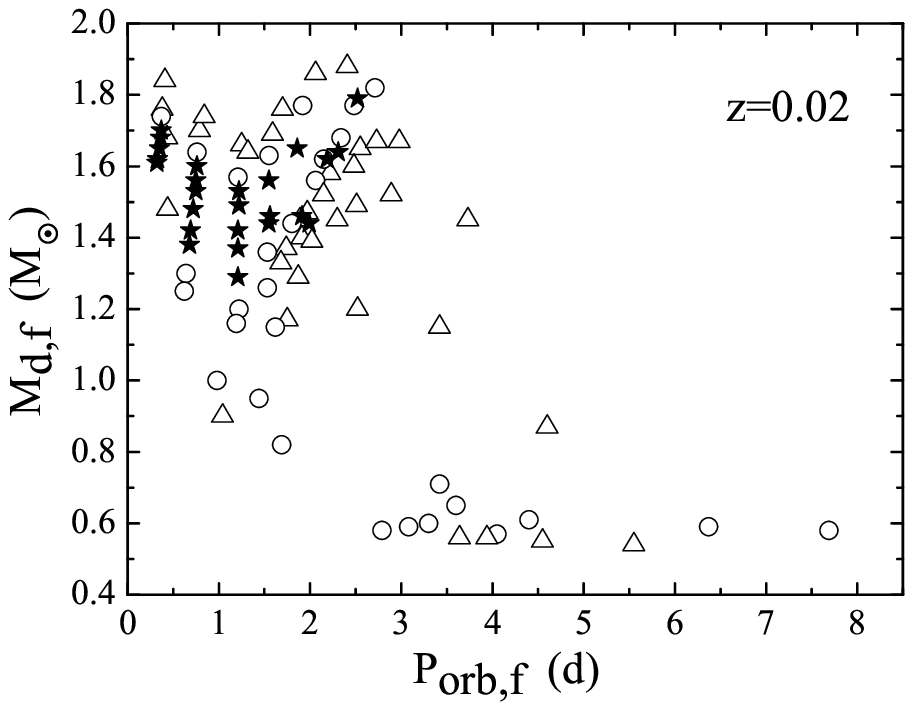}{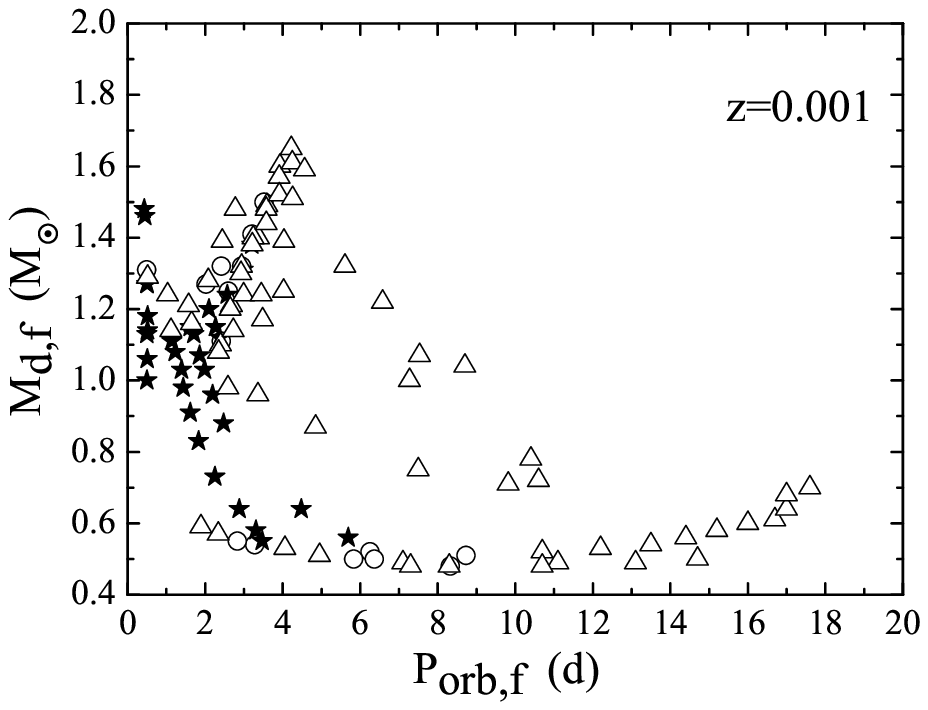}\caption{The donor star masses $M_{\rm
d,f}$ vs. the orbital periods $P_{\rm orb,f}$  when SN Ia
explosions occur. In the left panel ($Z=0.02$ ), the open
triangles, open circles, and solid stars denote systems with
$1.6M_{\odot}>M_{\rm SN}\ge 1.4M_{\odot}$, $1.7M_{\odot}>M_{\rm
SN}\ge 1.6M_{\odot}$, and $M_{\rm SN}\ge 1.7M_{\odot}$,
respectively. In the right panel ($Z=0.001$), the open triangles,
open circles, and solid stars denote systems with
$1.5M_{\odot}>M_{\rm SN}\ge 1.4M_{\odot}$, $1.6M_{\odot}>M_{\rm
SN}\ge 1.5M_{\odot}$, and $M_{\rm SN}\ge 1.6M_{\odot}$,
respectively. } \label{FigVibStab}
\end{figure}

\begin{figure}
\plottwo{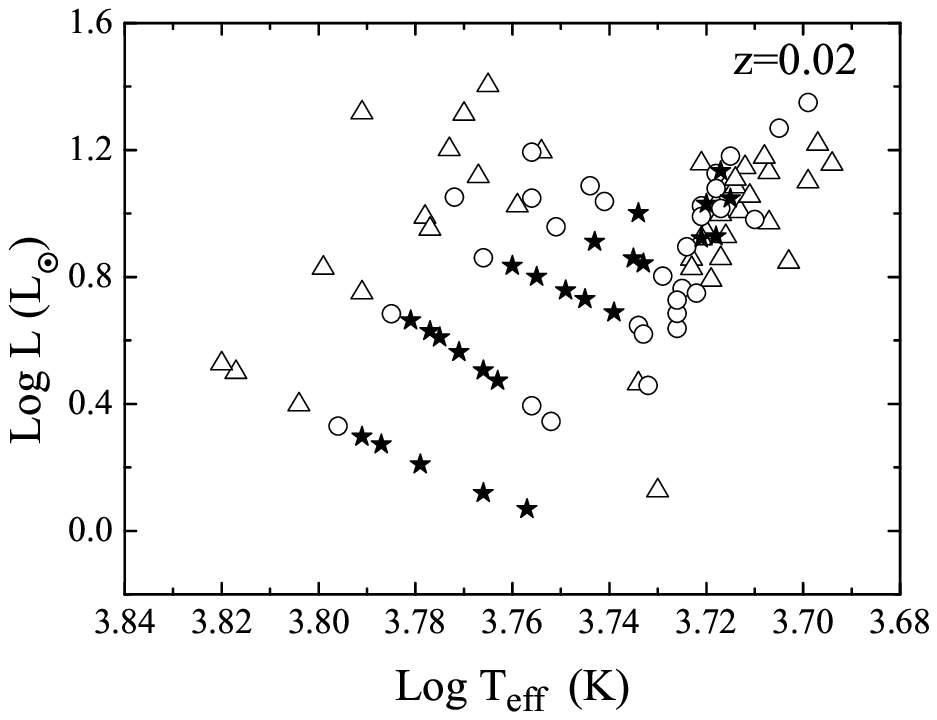}{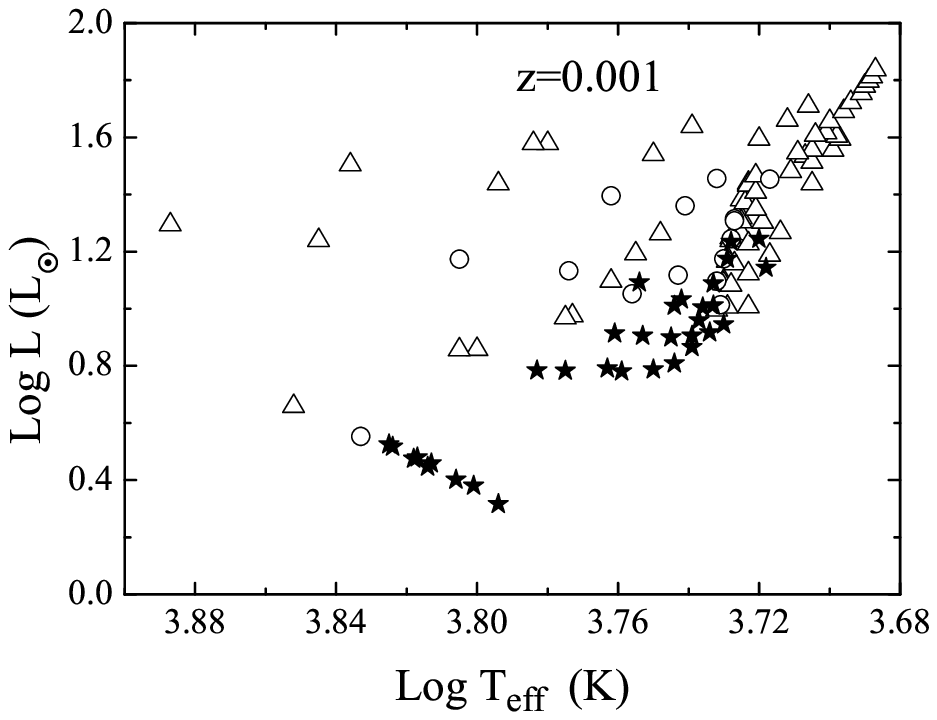}\caption{Distribution of the donor stars
in the H-R diagram when SN Ia explosions occurs. Symbols are the
same as in Figure 4. } \label{FigVibStab}
\end{figure}

\begin{figure}
\plotone{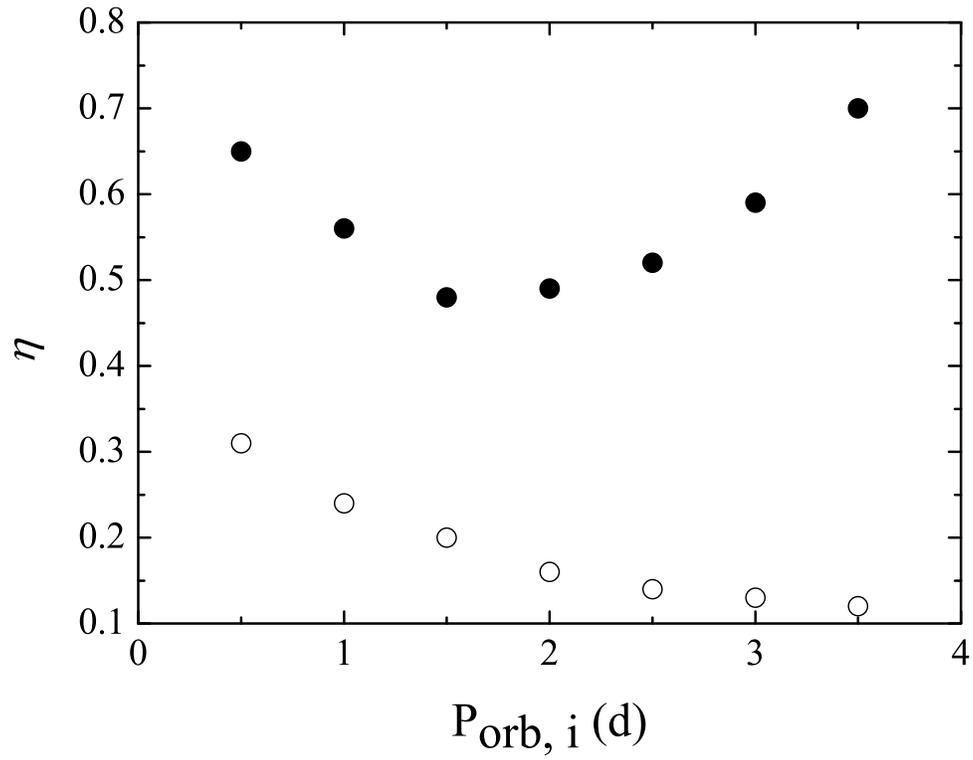} \caption{The  parameters $\eta$ vs. the initial
orbital periods $P_{\rm orb,i}$  when the initial masses of donor
stars $M_{\rm d.i}=2.5M_{\odot}$. The filled circles and open
circles correspond to the calculated results for $Z=0.02$ and
0.001, respectively. } \label{FigVibStab}
\end{figure}


\begin{thebibliography}{}
\bibitem[Alexander \& Ferguson (1994)]{alex94} Alexander D. R., Ferguson J. W., 1994, ApJ, 437, 879
\bibitem[Astier et al. (2006)]{asti06} Astier, P., et al. 2006, A\&A, 447, 31
\bibitem[Branch et al. (1995)]{bran95} Branch, D., Livio, M., Yungelson, L. R., Boffi, F. R., \& Baron, E. 1995, PASP, 107, 717
\bibitem[Chandrasekhar (1931)]{chan31} Chandrasekhar, S. 1931, ApJ, 74, 81
\bibitem[Chen \& Li (2007)]{chen07} Chen, W. -C., \& Li, X. -D. 2007, ApJ, 658, L51
\bibitem[Dom\'{i}nguez et al. (2006)]{Domi06} Dom\'{i}nguez, I., Piersanti, L., Bravo, E., Tornamb\'{e}, A., Straniero, O., \& Gagliardi, S. 2006, ApJ, 644, 21
\bibitem[Dom\'{i}nguez et al. (1996)]{Domi96} Dominguez, I., Straniero, O., Tornambe« , A., \& Isern, J. 1996, ApJ, 472, 783
\bibitem[Durisen (1975)]{Duri75} Durisen, R. H. 1975, ApJ, 199, 179
\bibitem[Durisen (1977)]{Duri77} Durisen, R. H. 1977, ApJ, 213, 145
\bibitem[Durisen \& Imamura (1981)]{Duri81} Durisen, R. H., \& Imamura, J. N. 1981, ApJ, 243, 612
\bibitem[Eggleton (1971)]{eggl71} Eggleton, P. P. 1971, MNRAS, 151, 351
\bibitem[Eggleton (1972)]{eggl72} Eggleton, P. P. 1972, MNRAS, 156, 361
\bibitem[Hachisu, Kato \& Nomoto (1996)]{hach96} Hachisu, I., Kato, M., \& Nomoto, K. 1996, ApJ, 470, L97
\bibitem[Hachisu et al. (1999)]{hach99b} Hachisu, I., Kato, M., Nomoto, K., \& Umeda, H. 1999, ApJ, 519, 314
\bibitem[Hachisu, Kato \& Nomoto (1999)]{hach99a} Hachisu, I., Kato, M., \& Nomoto, K. 1999, ApJ, 522, 487
\bibitem[Hachisu et al. (2008)]{hach08} Hachisu, I., Kato, M., \& Nomoto, K.  2008, ApJ, 679, 1390
\bibitem[Hillebrandt, Sim \& R\"{o}pke(2007)]{hill07} Hillebrandt, W., Sim, S. A., \& R\"{o}pke, F. K. 2007, A\&A, 465, L17
\bibitem[Han, Podsiadlowski \& Eggleton (1994)]{han94} Han, Z., Podsiadlowski, P., \& Eggleton, P. P. 1994, MNRAS, 270, 121
\bibitem[Han \& Podsiadlowski (2004)]{han04} Han, Z., \& Podsiadlowski, Ph. 2004, MNRAS, 350, 1301
\bibitem[Han (2008)]{han08} Han, Z. 2008, ApJ, 677, L109
\bibitem[Heger \& Langer (2000)]{hege00} Heger, A., \& Langer, N. 2000, ApJ, 544, 1016
\bibitem[Hicken et al. (2007)]{hick07} Hicken, M., et al. 2007, ApJ, 669, L17
\bibitem[Howell et al. (2006)]{howe06} Howell, D. A., Sullivan, M., Nugent, P. E., et al. 2006, Nat, 443, 308
\bibitem[Howell (2001)]{howe01} Howell, D. A. 2001, ApJ, 554, L193
\bibitem[Hoyle \& Fowler (1960)]{hoy60} Hoyle, F., \& Fowler, W. A. 1960, ApJ, 132, 565
\bibitem[Iben \& Tutukov (1984)]{iben84} Iben, I. Jr., \& Tutukov, A. V. 1984, ApJS, 54, 335
\bibitem[Kato \& Hachisu (2004)]{kato04}Kato, M., \& Hachisu, I. 2004, ApJ, 613, L219
\bibitem[Kovetz \& Prialnik (1994)]{kove94} Kovetz, A., \& Prialnik, D. 1994, ApJ, 424, 319
\bibitem[Langer et al. (2000)]{lang00} Langer, N., Deutschmann, A., Wellstein, S., \& H\"{o}flich, P. 2000, A\&A, 362, 1046
\bibitem[Li \& van den Heuvel(1997)]{li97} Li, X. -D., \& van den Heuvel, E. P. J. 1997, A\&A, 322, L9
\bibitem[Maeda et al. (2009)]{maed08a} Maeda, K., et al. 2009, ApJ, 690, 1745
\bibitem[Maeda \& Iwamoto (2009)]{maed08b} Maeda, K., \& Iwamoto, K. 2009, MNRAS, 394, 239
\bibitem[Maeder \& Meynet (2000)]{maed00} Maeder, A., \& Meynet, G. 2000, ARA\&A, 38, 143
\bibitem[Mazzzli et al. (2007)]{mazz07}Mazzali P. A., R\"{o}pke F. K., Benetti S., \& Hillebrandt, W. 2007, Science, 315, 825
\bibitem[Meng, Chen \& Han (2009)]{meng08} Meng, X., Chen, X., \& Han, Z. 2009, MNRAS, 395, 2103
\bibitem[Narayan \& Pophm (1989)]{nara89}Narayan, R., \& Popham, R. 1989, ApJ, 346, L25
\bibitem[Nomoto (1982)]{nomo82} Nomoto, K. 1982, ApJ, 253, 798
\bibitem[Paczy\'{n}ski (1991)]{pacz91} Paczy\'{n}ski, B. 1991, ApJ, 370, 597
\bibitem[Perlmutter et al. (1999)]{perl99} Perlmutter, S., et al. 1999, \apj, 517, 565
\bibitem[Phillips (1993)]{phil93} Phillips, M. M. 1993, ApJ, 413, L105
\bibitem[Piersanti et al. (2003)]{pier03a}Piersanti, L., Gagliardi, S., Iben, I.Jr., \& Tornambe, A., 2003, ApJ, 583, 885
\bibitem[Pols et al.(1995)]{pols95} Pols, O., Tout, C. A., Eggleton, P. P., \& Han, Z. 1995, MNRAS, 274, 964
\bibitem[Popham, \& Narayan (1991)]{poph91} Popham, R., \& Narayan, R. 1991, ApJ, 370, 604
\bibitem[Riess et al. (1998)]{ries98} Riess, A. G. et al. 1998, AJ, 116, 1009
\bibitem[Riess et al. (2004)]{ries04} Riess, A. G. et al. 2004, \apj, 607, 665
\bibitem[Ritter (1985)]{ritt85} Ritter, H. 1985, A\&A, 148, 207
\bibitem[Rogers \& Iglesias (1992)]{roge92} Rogers F. J., Iglesias C. A., 1992, ApJS, 79, 507
\bibitem[Saio \& Nomoto (2004)]{saio04}Saio, H., \& Nomoto, K.I., 2004, ApJ, 615, 444
\bibitem[Shapiro \& Teukolsky (1983)]{shap83} Shapiro, S. L., \& Teukolsky, S. A. 1983, Black holes, white dwarfs, and neutron stars: The physics of
compact objects (Wiley-Interscience, New York)
\bibitem[Tutukov \& Yungelson (1994)]{tutu94} Tutukov, A. V., \& Yungelson, L. R. 1994, MNRAS, 268, 871
\bibitem[Uenishi et al. (2003)]{ueni03}Uenishi, T., Nomoto, K.I., Hachisu, I., 2003, ApJ, 595, 1094
\bibitem[Wang et al. (2008)]{wang08} Wang, B., Meng, X. C., Wang, X. F., \& Han, Z. W. 2008, ChJAA, 8, 71
\bibitem[Wang et al. (2009)]{wang09} Wang, B., Meng, X. C., Chen, X., Han, Z. W. 2009, MNRAS, 395, 847
\bibitem[Webbink (1984)]{webb84} Webbink, R. F. 1984, ApJ, 277, 355
\bibitem[Whelan \& Iben (1973)]{whel73} Whelan, J., \& Iban, I. Jr. 1973, ApJ, 186, 1007
\bibitem[Xu \& Li (2009)]{xu08} Xu, X.-J., \& Li, X.-D. 2009, A\&A, 495, 243
\bibitem[Yoon \& Langer (2003)]{yoon03} Yoon, S. -C., \& Langer, N.  2003, A\&A, 412, L53
\bibitem[Yoon \& Langer (2004)]{yoon04a} Yoon, S. -C., \& Langer, N.  2004, A\&A, 419, 623
\bibitem[Yoon, Langer \& Scheithauer (2004)]{yoon04b} Yoon, S. -C., Langer, N., \& Scheithauer, S.  2004, A\&A, 425, 217
\bibitem[Yoon \& Langer (2005)]{yoon05} Yoon, S. -C., \& Langer, N.  2005, A\&A, 435, 967
\bibitem[Yuan et al. (2007)]{yuan07} Yuan, R. F. et al. 2007, ATEL, 1212

\end{thebibliography}
\end{document}